\documentclass[final,5p,times,twocolumn]{elsarticle}

\usepackage{alltt}
\usepackage{setspace}
\usepackage{color}
\usepackage{bbm}
\usepackage{mathtools}
\usepackage{amsmath,amsfonts,amssymb,txfonts}
\usepackage{mathrsfs}
\usepackage{graphicx}% Include figure files
\usepackage{dcolumn}% Align table columns on decimal point
\usepackage{bm,bbold}% bold math
\usepackage{comment,xcolor,slashed,empheq}% 
\usepackage[colorlinks=true,breaklinks=true]{hyperref}
\hypersetup{allcolors=[rgb]{0.0 0.0 0.6},linkcolor=[rgb]{0.75 0.05 0.05}}

\usepackage[capitalise]{cleveref}
\crefformat{pluralequation}{#2{\color{black} Eqs.~(}#1{\color{black} )}#3}
\Crefformat{pluralequation}
{#2{\color{black} Equations~(}#1{\color{black} )}#3}
\def\im	{\mathbb{i}}
\newcommand\widefbox[1]{\fbox{\hspace{1ex}#1\hspace{1ex}}}

\journal{Physics Letters B}

\begin{document}

\begin{frontmatter}

\title{Unitarity in the non-relativistic regime and implications for dark matter}

\author{Marcos M. Flores}
\ead{marcos.flores@phys.ens.fr}

\author{Kalliopi Petraki}
\ead{kalliopi.petraki@phys.ens.fr}

\address{Laboratoire de Physique de l'École Normale Supérieure, ENS, Université PSL, CNRS, Sorbonne Université, Université Paris Cité, Paris, F-75005, France}

\begin{abstract}
Unitarity sets upper limits on partial-wave elastic and inelastic cross-sections, which are often violated by perturbative computations. We discuss the dynamics underlying these limits in the non-relativistic regime, namely long-range interactions, and show how the resummation of the 2-particle-irreducible diagrams arising from squaring inelastic processes unitarizes both elastic and inelastic cross-sections. We provide a simple prescription to obtain the unitarized cross-sections from those that do not include resummation of the squared inelastic processes. Our results are model-independent, apply to all partial waves, and affect elastic and inelastic cross-sections, with extensive implications for new physics scenarios, such as dark-matter freeze-out, indirect detection and self-interactions.
\end{abstract}


\end{frontmatter}

%%% main text %%%
\section{Introduction \label{sec:Intro} }

Unitarity has long stood as a fundamental principle in particle physics,  crucially invoked in the past to argue for the existence of then-unknown physics, including the electroweak interactions and the Higgs mechanism~\citep{Weinberg:1971fb,Cornwall:1974km,Lee:1977yc}. More recently, it has been extensively employed to circumscribe the range of validity of effective theories aiming to parametrize the still-unknown physics beyond the Standard Model. In this respect, the implications of unitarity have been widely considered in the high-energy limit, relevant for new physics searches at colliders.

In this letter, we consider unitarity in the antipodal limit, the non-relativistic regime~\citep{Baldes:2017gzw}. In the forefront of our motivation lies dark matter (DM), which makes up 85\% of the non-relativistic matter in our Universe. Besides the DM phenomenology in today's Universe, many important cosmological events, such as the DM production in the early universe, may also fall within the non-relativistic realm. Our results pertain to the cosmology and astrophysics of DM in a variety of theories at the frontiers of research, but are also more generally applicable to any non-relativistic system. 

Unitarity sets upper bounds on partial-wave elastic and inelastic cross-sections that indicate the saturation of the corresponding scattering probability to one. In a theory that is by construction unitary, the ramifications of these bounds are two-fold: 
($a$)~They may limit unknown physical parameters if phenomenological requirements are imposed that amount to certain processes being sufficiently efficient.  
($b$)~They indicate the point where approximations employed to compute scattering amplitudes fail. While this failure may be simply due to the break-down of a perturbative expansion in a parameter that has become large, in certain cases it hints towards the onset of new effects.

In the present work, we reframe these points, posing the questions: 
($a'$) What theories give rise to physical processes that can approach or attain the unitarity limits on scattering cross-sections, thus allowing to approach or attain the bounds on physical parameters set by phenomenological requirements? ($b'$) How can we improve perturbative calculations in weakly-coupled unitary theories to ensure consistency with unitarity? We begin with reviewing the unitarity bounds, before addressing these points and discussing examples in the DM context.

\section{Off-shell partial-wave optical theorem \label{sec:OpticalTheorem}}

To minimize technicalities that are tangential to the main goal of this work, we neglect spin in the following. We consider 2-to-2 transition amplitudes that may in general be \emph{off-shell}, an essential ingredient in our analysis. We expand them into partial waves as follows,
\begin{multline}
{\cal M}^{ab}(s, {\bf k}^a,{\bf k}^b) = \\
= 16\pi \sum_{\ell} 
(2\ell+1) P_{\ell}({\bf \hat{k}}^a \cdot {\bf \hat{k}}^b) 
{\cal M}^{ab}_{\ell} (s,|{\bf k}^a|,|{\bf k}^b|) ,
\label{eq:PartialWaveExpansion_Amplitude}
\end{multline}
where $a$ and $b$ denote the initial and final state, $s$ is the first Mandelstam variable, and ${\bf k}_{a}$ stands for the 3-momentum of each particle in the center-of-momentum (CM) frame in the state $a$. Note that the two interacting particles may in general be different. Next, we define the rescaled partial-wave amplitude
\begin{align}
M^{ab}_{\ell} (s,|{\bf k}^a|,|{\bf k}^b|) \equiv 
\sqrt{\dfrac{4|{\bf k}^a||{\bf k}^b|}{2^{\delta_a}2^{\delta_b} s}}
{\cal M}^{ab}_{\ell} (s,|{\bf k}^a|,|{\bf k}^b|) ,
\label{eq:Amplitudes_QMvsQFT_def}
\end{align}
where we have introduced the symmetry parameter $\delta_a = 0$~or~1 if the two particles in the state $a$ are distinguishable or identical respectively.

Following standard methods generalized to off-shell amplitudes~\citep{Kowalski:1966zz}, $S$-matrix unitarity implies
\begin{align}
{\rm Im} [M^{aa}_{\ell} (s,|{\bf k}|,|{\bf k'}|)] 
=\sum_{b:~\text{on-shell}} X_\ell^{ab} (s,|{\bf k}|,|{\bf k'}|),
\label{eq:OpticalTheorem_gen}
\end{align}
where $b$ runs over all on-shell states into which $a$ can scatter, and
\begin{align}
X_\ell^{ab} (s,|{\bf k}|,|{\bf k'}|) &\equiv    
\sqrt{\dfrac{4|{\bf k}| |{\bf k'}|}{2^{2\delta_a}s}}
\times
\dfrac{1}{2\,(8\pi)^2} 
\int d\Pi_b \, (2\pi)^4 \delta^4 (\{k\} - \{k^b\}) 
\nonumber 
\\
&\times \left[
\int d\Omega_{\bf k} \, Y_{\ell m}^*(\Omega_{\bf k})
{\cal M}^{ab} (s,{\bf k},\{{\bf k}^b\})
\right]
\nonumber 
\\
&\times \left[
\int d\Omega_{\bf k'} \, Y_{\ell m} (\Omega_{\bf k'})
{\cal M}^{ab*}  (s,{\bf k'},\{{\bf k}^b\}) 
\right].
\label{eq:OpticalTheorem_IndividualContributions}
\end{align}
\Cref{eq:OpticalTheorem_IndividualContributions} holds for any $m \in [-\ell,\ell]$, and can also be averaged over $m$. 
Curly brackets denote collectively the momenta of multi-particle states, and $d\Pi_b$ is the phase-space integration measure for the $n$-particle state $b$, 
%multi
\begin{align}
d\Pi_b = f_b \prod_{j=1}^n 
\dfrac{d^3k_j^b}{(2\pi)^3 \, 2E_j^b}  ,
\label{eq:PhaseSpaceIntegrationMeasure}
\end{align}
where $f_b$ is the symmetry factor of the state $b$ that ensures the phase space is not multiply counted. For 2-particle states $b$, $f_b=1/2^{\delta_b}$, and \cref{eq:OpticalTheorem_IndividualContributions} reduces to
\begin{align}
\label{eq:OpticalTheorem_2-ParticleContribution}
X_\ell^{ab} (s,|{\bf k}|,|{\bf k'}|) = 
M^{ab \, *}_\ell (s,|{\bf k}|,|{\bf k}^b|) \, 
M^{ab}_{\ell} (s,|{\bf k'}|,|{\bf k}^b|) ,
\end{align}
with $|{\bf k}^b|$ determined by $s$ due to $b$ being on-shell. 
\Cref{eq:OpticalTheorem_gen,eq:OpticalTheorem_IndividualContributions,eq:PhaseSpaceIntegrationMeasure,eq:OpticalTheorem_2-ParticleContribution} generalize the optical theorem, and we shall employ them in different ways below.

Notably, $X_\ell^{ab} (s,|{\bf k}|,|{\bf k}|)  \geqslant 0$, which leads to the important observation that the sign of the imaginary part of the elastic amplitude is not arbitrary, ${\rm Im} [M^{aa}_{\ell}(s,|{\bf k}|,|{\bf k}|)] \geqslant 0$. This fact will prove critical in the following.

\section{Unitarity bounds \label{sec:UnitarityBounds}}

%%%%%%%%%%%%%%%%%%%%%%%%%%%
\begin{figure}[t!]
\centering
\includegraphics[width=0.95\linewidth] {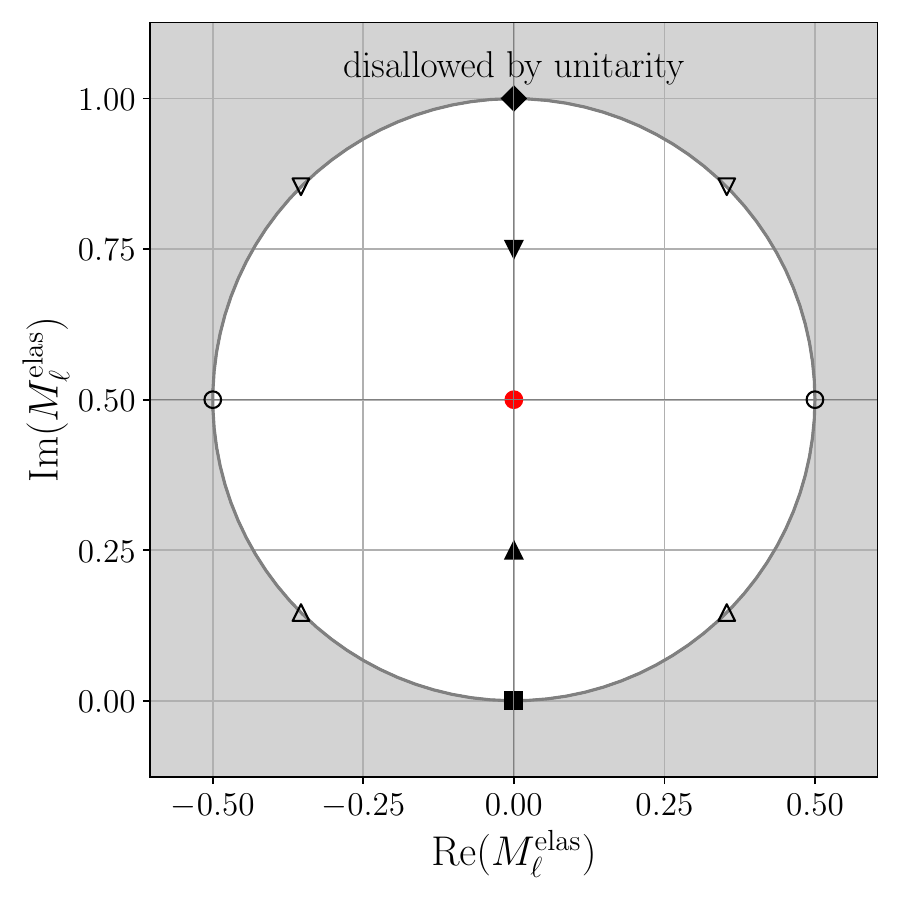} \\[1em]
\includegraphics[width=0.95\linewidth] {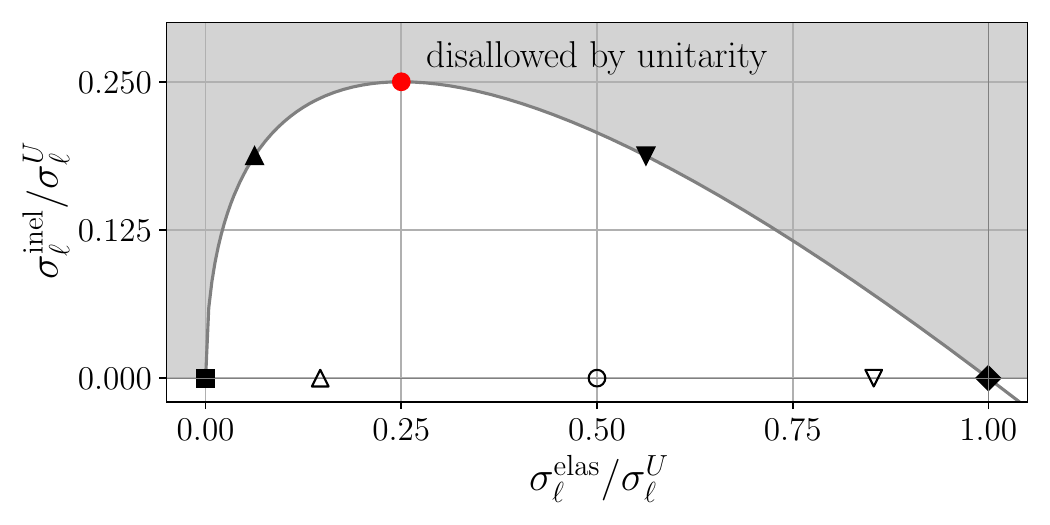}
\caption{
Upper: The unitarity circle bounding the on-shell partial-wave elastic matrix elements $M^{\rm elas}_\ell$. The border of the circle corresponds to vanishing inelasticity. 
Lower: Combined bound on the partial-wave elastic and inelastic cross-sections. The symbols identify points between the two plots. The filled red circle corresponds to maximum inelastic scattering.
}
\label{fig:Unitarity}
\end{figure}
%%%%%%%%%%%%%%%%%%%%%%%%%%%

Next, we focus on \emph{on-shell} in and out states. We define the cross-section
\begin{empheq}[box=\widefbox]{align} 
\sigma^U_\ell \equiv 
2^\delta 4\pi (2\ell+1) / {\bf k}^2,
\label{eq:sigmaU}
\end{empheq}
with the symmetry parameter $\delta$ and the CM momentum ${\bf k}$ referring to the incoming state. Then, all cross-sections can be expressed in the convenient form 
\begin{align}
\sigma_\ell^{a\to b} = 
\sigma_\ell^U |M_\ell^{ab}|^2,  
\label{eq:sigma2to2_wrt_rescaledM}
\end{align} 
where for 2-particle states, $M_\ell^{ab}$ is given by \cref{eq:Amplitudes_QMvsQFT_def}, while for states of different multiplicity we set $|M_\ell^{ab}|^2 \to X_\ell^{ab} (s,|{\bf k}|,|{\bf k}|)$, with $|{\bf k}|$ fixed by $s$ due to the incoming state being on-shell.

For on-shell states, \cref{eq:OpticalTheorem_gen} implies
\begin{align}
{\rm Im}(M^{aa}_\ell) 
= \sum_b |M^{ab}_\ell|^2
\geqslant |M^{aa}_\ell|^2 .
\label{eq:UnitarityBounds_Amplitudes}
\end{align}
This can be recast for $M^{\rm elas}_\ell = M^{aa}_\ell$ as 
\begin{align}
|M_\ell^{\rm elas} - \im/2| \leqslant 1/2, 
\label{eq:UnitarityCircle}
\end{align}
which is the familiar unitarity circle centered at $\im/2$ on the complex plane. 
We may obtain a more restrictive combined bound on elastic and inelastic cross-sections as follows. The inequality \cref{eq:UnitarityCircle} implies ${\rm Im (M_\ell^{\rm elas})} \leqslant |M_\ell^{\rm elas}|$. Then, recalling \cref{eq:sigma2to2_wrt_rescaledM}, 
\begin{align*}
(\sigma_\ell^{\rm elas}+\sigma_\ell^{\rm inel}) / \sigma_\ell^U
&= |M_\ell^{\rm elas}|^2 + \sum_j |M_\ell^{{\rm inel},j}|^2
\\ 
&=  {\rm Im} |M_\ell^{\rm elas}| 
\leqslant  |M_\ell^{\rm elas}| 
=  (\sigma_{\ell}^{\rm elas} / \sigma_\ell^U)^{1/2} ,
\end{align*}
which can be recast as
\begin{empheq}[box=\widefbox]{align} 
\sigma^{\rm inel}_\ell /  \sigma^U_\ell \leqslant
\sqrt{\sigma^{\rm elas}_\ell /  \sigma^U_\ell}
\left(1-\sqrt{\sigma^{\rm elas}_\ell /  \sigma^U_\ell} \right).
\label{eq:UnitarityBound_CrossSections_Combined}
\end{empheq}
The above encompasses the individual bounds 
$\sigma^{\rm inel}_\ell \leqslant  \sigma^U_\ell/4$ and
$\sigma^{\rm elas}_\ell \leqslant  \sigma^U_\ell$.
It agrees with~\citep{Hui:2001wy} except for the symmetry factor $2^\delta$ in $\sigma^U_\ell$ that has been missed in the past. 
The unitarity circle \eqref{eq:UnitarityCircle}, the combined bound on elastic and inelastic cross-sections \eqref{eq:UnitarityBound_CrossSections_Combined}, and the mapping between them are depicted in \cref{fig:Unitarity}.

\section{The physics of the unitarity limit \label{sec:UnitarityLimitInterpretation}}

\subsection{Attaining the unitarity limit
\label{sec:UnitarityLimitInterpretation_Dynamics}}

The dependence of the cross-section~\eqref{eq:sigmaU} on the kinematic variables allows us to address question $(a')$. Let us first motivate this phenomenologically with an example. 

Reference~\citep{Griest:1989wd} argued that the inelastic unitarity bounds imply an upper bound on the mass of thermal-relic DM. The calculation amounted to bounding the so-called canonical DM annihilation cross-section by the inelastic $s$-wave unitarity limit, 
$\sigma_{\rm can} v_{\rm rel} = 
2.2\times 10^{-26}{\rm cm^3/s} < 
4\pi / (m_{\rm DM}^2 v_{\rm rel})$, where we set $|{\bf k}| = m_{\rm DM} v_{\rm rel}/2$ in the non-relativistic regime, with $m_{\rm DM}$ and $v_{\rm rel}$ being the mass and relative velocity of the annihilating DM particles, and we used the current measurement of the DM density to determine $\sigma_{\rm can}$. In this comparison, there is clearly a discrepancy between the velocity scalings of $\sigma_{\rm can}$ and $\sigma_0^U$. Reference~\citep{Griest:1989wd} claimed that the momentum scaling of $\sigma_\ell^U$ cannot be obtained in the non-relativistic regime (and simply set $v_{\rm rel}^{\rm freeze-out} \sim 0.5$ in order to evaluate $\sigma^U_0$), and that $s$-wave annihilation dominates. This however, is not so, as unitarity implies~\citep{Baldes:2017gzw,vonHarling:2014kha} and we shall now discuss. This, in turn, has implications for the upper limit $m_{\rm DM} < 116$~ (82)~TeV, deduced according to the considerations of Ref.~\cite{Griest:1989wd}\footnote{We note that the computation of Ref.~\cite{Griest:1989wd} of the upper limit on $m_{\rm DM}$ is in conflict with unitarity. The DM particles are presumed to have a velocity distribution in the early universe. Setting $\sigma_{\rm can}$ equal to the $s$-wave inelastic unitarity limit evaluated at a fixed $v_{\rm rel}$ implies that particles with higher velocity violate unitarity, while particles with lower velocity do not saturate their unitarity limit.} 
for (non-)self-conjugate thermal-relic DM, but, more importantly, for the underlying physics of viable thermal-relic DM scenarios at the multi-TeV mass scale~\citep{Baldes:2017gzw}.

In the relativistic regime, $|{\bf k}| \to \sqrt{s}/2$, thus $\sigma_\ell^U \simeq 2^{\delta} 16\pi(2\ell+1)/s$, leading to the well-known conclusion that cross-sections should decrease at least as fast as $\sigma \propto1/s$ at high energies, in order to be consistent with unitarity up to arbitrary high scales. However, for processes computed within an effective theory where  a large scale $\Lambda>m$ has been integrated out, with $m$ being the mass of the interacting particles, the cross-sections scale as $\sigma \propto (1/s) (s/\Lambda^2)^p$, where $p>0$ depends on the effective theory. Unitarity then implies that the effective theory breaks down at $s\gtrsim \Lambda^2$, where the high-energy physics must be resolved and incorporated in the computation. 
If doing so realizes the scaling $\sigma \propto 1/s$,  then approaching or attaining the unitarity limit in a continuum of energies $s> \Lambda^2$ amounts essentially to the couplings involved being sufficiently large.

If $\Lambda \lesssim m$, then the breakdown of the effective theory and the transition described above occur in the quasi-relativistic or non-relativistic regime. As previously, the scaling $\sigma \propto 1/s$ can retain consistency with unitarity at increasing energies since $s > |{\bf k}|^2$, thus $\sigma_\ell^U \propto 1/|{\bf k}|^2 > 1/s$. However, now, this scaling does not allow to approach or attain the unitarity limit in a continuum of low momenta, $|{\bf k}| \lesssim m$. To identify what can realize the unitarity limit in this case, we repeat the previous considerations in the non-relativistic regime.

For concreteness, we consider a toy model, and follow in part the discussion of Refs.~\cite{Hisano:2002fk,Baldes:2017gzw}. Similar considerations to what follows can be employed in a variety of theories, including gauge theories. Let $\chi$ be a scalar of mass $m$, coupled to a scalar force mediator $\varphi$ of mass $M$ via the trilinear coupling ${\cal L} \supset - (y/2) m \, \chi^2 \varphi$, with $y$ being a dimensionless coupling. We define $\alpha \equiv y^2/(16\pi)$ for convenience. The mediator $\varphi$ may also couple in a similar fashion to other lighter species. $\chi\chi$ pairs can scatter elastically or annihilate via $\varphi$ exchanges. If $\varphi$ is sufficiently heavy, then it can be effectively integrated out, giving rise to the $\chi\chi$ elastic and annihilation cross-sections $\sigma^{\rm elas} \sim \sigma^{\rm ann} v_{\rm rel} \sim \alpha^2 s/M^4$, with $v_{\rm rel}$ being the $\chi\chi$ relative velocity. Setting $s \simeq 4m^2$ and $|{\bf k}|\simeq m v_{\rm rel}/2$, unitarity implies, as previously, that this effective description breaks down at sufficiently high energies -- with the energy now more conveniently parametrized by $|{\bf k}|$ or $v_{\rm rel}$ -- in particular at $|{\bf k}| > M^2/(m\alpha)$. This can be recast in the more indicative form, $M < m \sqrt{\alpha v_{\rm rel}} \ll m$. 

The comparison between scales that arises from this consideration hints to the reason for the apparent violation of unitarity: $\varphi$ mediates a $\chi\chi$ interaction of range $\lambda_{\rm inter} \sim M^{-1}$ that is comparable to or longer than the geometric mean of the de Broglie wavelength of the interacting particles, $\lambda_{dB} \sim |{\bf k}|^{-1} = (m v_{\rm rel}/2)^{-1}$ and their Bohr radius, $\lambda_{B} \sim (m\alpha/2)^{-1}$. Consequently, virtual $\varphi$ exchanges cause \emph{long-range} elastic $\chi\chi$ interactions that distort the wavefunction of the incoming $\chi\chi$ pair, and must thus be \emph{resummed} in order to properly compute both the elastic and inelastic cross-sections. (More details on resummation will be provided in the next section.) In the context of inelastic scattering, this is called the Sommerfeld effect~\cite{Sakharov:1948yq,Sommerfeld:1931}. As we discuss next, the resummation of an (attractive) long-range interaction reproduces the same energy dependence as that of the unitarity cross-section~\eqref{eq:sigmaU}.

The resummation of the $t/u$-channel $\varphi$ exchanges generates an attractive Yukawa potential between the $\chi$ particles,  $V(r) = -\alpha e^{-Mr}/r$ (see e.g.~\cite{Petraki:2015hla}). In the Coulomb limit (which is valid for $M < m v_{\rm rel}/2$, see~e.g.~\cite{Petraki:2016cnz}), this results in the cross-sections 
\begin{subequations}
\label{eq:LongRangeScattering}
\label[pluralequation]{eqs:LongRangeScattering}
\begin{align}
\sigma_\ell^{\rm elas} &=  2\dfrac{4\pi(2\ell+1)}{\bf k^2}
\sin^2 \left(\dfrac{1}{2\im}
\ln\left[\dfrac{\Gamma(1+\ell+\im \zeta)}{\Gamma(1+\ell-\im \zeta)}\right]
\right),
\label{eq:LongRangeScattering_Elas}
\\
\sigma_{\ell}^{\rm ann} v_{\rm rel} 
&\sim \dfrac{\pi\alpha^2}{m^2} v_{\rm rel}^{2\ell} 
\ S_\ell(\zeta),
\label{eq:LongRangeScattering_Ann}
\end{align}
\end{subequations}
where $\zeta \equiv \alpha / v_{\rm rel}$. $S_\ell (\zeta) = S_0(\zeta) \prod_{j=1}^\ell (1+\zeta^2/j^2)$ with  $S_0(\zeta) = 2\pi \zeta / (1-e^{-2\pi\zeta})$ are the Sommerfeld factors for a Coulomb potential. 
Expanding for small values of the coupling, $\zeta \ll 1+\ell$, the above reduce to the well-known Rutherford scattering,
$\sigma_\ell^{\rm elas} \propto \alpha^2/(m^2 v_{\rm rel}^4)$, and the tree-level annhilation cross-sections, $\sigma^{\rm ann}_{\ell} v_{\rm rel} \sim (\pi\alpha^2/m^2) v_{\rm rel}^{2\ell}$. 
On the other hand, for large values of the coupling, 
$\zeta \gg 1+\ell$, 
$\sigma^{\rm ann}_{\ell} \propto \alpha^{2\ell+3} \sigma_\ell^U$ while the elastic cross-section oscillates rapidly in the interval 
$\sigma_\ell^{\rm elas} \in [0,\sigma_\ell^U]$ with a $\zeta$-dependent phase. The annihilation cross-sections thus scale with momentum in the same way as the unitarity limit,\footnote{In \cref{eq:LongRangeScattering_Ann}, we have considered only the leading order contribution in $v_{\rm rel}^2$ for every partial wave. Higher order contributions receive different Sommerfeld corrections that result in the same asymptotic behavior at large $\alpha/v_{\rm rel}$~\cite{ElHedri:2016onc,Baldes:2017gzw}.} which they may approach or attain in a continuum of momenta provided that the coupling is sufficiently large. Conversely, we conclude that the inelastic unitarity limit in the non-relativistic regime can be realized in a continuum of energies only if the incoming particles interact via a long-range force~\citep{Baldes:2017gzw,vonHarling:2014kha}. Indeed, an interaction of range comparable to or smaller than $|{\bf k}|^{-1}$ introduces a new scale -- the range of the interaction -- that can only spoil the $1/|{\bf k}|^2$ scaling of the cross-sections.\footnote{It is reasonable to wonder whether other, perhaps exotic, interactions could result in cross-sections that scale as the unitarity limit in the non-relativistic regime.  Any exotic mechanism that could generate a Coulomb potential would be encompassed by what we call a long-range force. On the other hand, a different potential cannot result in Coulomb wavefunctions, which is what gives rise to cross-sections with the same momentum dependence as the unitarity limit. This reinforces the above statement. Nevertheless, the unitarity limit may be approached or attained by other types of interactions  within a limited energy range, e.g.~interactions of finite range at resonant points.
}

Returning to the phenomenological example of DM freeze-out, we may conclude that thermal-relic DM can be as heavy as unitarity allows only if it possesses attractive long-range interactions~\citep{Baldes:2017gzw}. Such interactions imply the existence of bound states, and it has been shown that the formation and decay of metastable bound states in the early Universe can deplete DM very efficiently~\citep{vonHarling:2014kha}, also via higher partial waves~\citep{Baldes:2017gzw}. 
The significant contribution of higher partial waves is in contrast to the direct annihilation processes, for which higher $\ell$ modes are suppressed by $v_{\rm rel}^{2\ell}$ or $\alpha^{2\ell}$ at tree level and in the Sommerfeld-enhanced regime respectively, as seen above. In the case of bound-state formation, the bound-state wavefunction introduces negative powers of the coupling, compensating for this suppression~\cite{Oncala:2019yvj,Binder:2023ckj}.
How many and which partial waves are important depends on the model, there is thus no model-independent upper bound on the mass of thermal-relic DM. Nevertheless, considering carefully the unitarity limit has allowed us to understand and predict, in a model-independent fashion, the underlying physics of viable heavy thermal-relic DM models.

\subsection{Unitarity violation \label{sec:UnitarityLimitInterpretation_Violation b}}

Our discussion of the dynamics that can approach or attain the (inelastic) unitarity limit in a continuum of energies in the non-relativistic regime, reframes the issue of (apparent) unitarity violation. In the toy model of \cref{sec:UnitarityLimitInterpretation_Dynamics}, the elastic cross-section \eqref{eq:LongRangeScattering_Elas} always respects the global limit $\sigma_\ell^{\rm elas} \leqslant \sigma_\ell^U$. However, the inelastic cross-sections \eqref{eq:LongRangeScattering_Ann} violate     
$\sigma_\ell^{\rm inel} \leqslant \sigma_\ell^U/4$ at large $\alpha$. Moreover the cross-sections \eqref{eq:LongRangeScattering} violate the combined unitarity bound \eqref{eq:UnitarityBound_CrossSections_Combined}. This is made apparent, for example, in the limit $\sigma_\ell^{\rm ann} \to \sigma_\ell^U/4$, where the elastic cross-section oscillates in the interval $[0,\sigma_\ell^U]$, in contradiction with the constraint \eqref{eq:UnitarityBound_CrossSections_Combined} which mandates that $\sigma_\ell^{\rm elas}$ must approach $\sigma_\ell^U/4$ as well (cf.~\cref{fig:Unitarity}).
This example represents one of several qualitatively distinct cases of unitarity violation observed in state-of-the-art computations in theories involving long-range interactions: 
\begin{enumerate}[(i)]
\item 
Attractive Coulomb(-like) potentials violate inelastic unitarity at large couplings. In a QED-like theory, this occurs at $\alpha \gtrsim 0.85$, considering the dominant $s$- and $p$-wave inelastic processes for a fermion-antifermion pair~\citep{Petraki:2016cnz}.  

\item
Radiative transitions between states of different Coulomb(-like) potentials, e.g.~bound-state formation with emission of a charged scalar~\citep{Oncala:2019yvj,Oncala:2021swy,Oncala:2021tkz,Ko:2019wxq} or a non-Abelian gauge boson~\citep{Harz:2018csl,Binder:2023ckj}, exhibit non-monotonic dependence on the incoming momentum, with peaks that can violate unitarity for a finite velocity range, even for fairly small values of the couplings. The problem is exacerbated if the potential in the scattering state is repulsive~\citep{Oncala:2019yvj,Binder:2023ckj}. 

\item
Potentials generated by the exchange of light but massive mediators exhibit parametric resonances at the thresholds for the existence of bound levels, violating inelastic unitarity at low $v_{\rm rel}$, even for small couplings~\citep{Cassel:2009wt,Petraki:2016cnz}. 

\end{enumerate}

These issues highlight a limitation in our understanding of this regime; they also hamper phenomenological studies of, for example, heavy thermal-relic DM or self-interacting DM, both of which feature long-range interactions. This brings us to question $(b')$ that we now address.

The inequalities \eqref{eq:UnitarityBounds_Amplitudes} cannot be satisfied if the amplitudes are calculated at any finite order in perturbation theory, as the two sides would then be of different order in the couplings of the theory. This suggests that in order to ensure consistency with unitarity, we must resum \emph{all} interactions of interest.
Indeed, returning to the cross-sections \eqref{eq:LongRangeScattering}, the resummation of the elastic $\chi\chi$ interaction (the one-boson exchange), resulted in the elastic cross-section \eqref{eq:LongRangeScattering_Elas} being consistent with the elastic unitarity bound for all momenta and values of the coupling. 
On the other hand, the inelastic cross-sections \eqref{eq:LongRangeScattering_Ann}, despite including the resummed one-boson exchanges, also involve a purely inelastic interaction that has not been resummed. However, the optical theorem implies that squaring inelastic interactions generates a contribution to elastic scattering; this must be self-consistently resummed in any computation that invokes these inelastic interactions, in order to ensure consistency with unitarity.

In the following we shall show that the proper resummation of the squared inelastic interactions indeed renders elastic and inelastic cross-sections consistent with the combined unitarity limit \eqref{eq:UnitarityBound_CrossSections_Combined}.  We begin with reviewing the resummation process, establishing the appropriate framework.

\section{Resummation \label{sec:Resummation}}

The resummation of 2-particle interactions amounts to a Dyson-Schwinger equation for the 4-point function of the two interacting particles, with the kernel being the sum of the 2-particle-irreducible (2PI) diagrams. The latter are defined as those that cannot be cut to give two individual contributions to the same 4-point function. The Dyson-Schwinger equation yields the Bethe-Salpeter equation for the two-particle wavefunction, which reduces to the Schr\"odinger equation under the instantaneous and non-relativistic approximations (see e.g.~\cite{Petraki:2015hla} for precise definitions and derivations). In momentum space, it reads
\begin{align}
&\left(\dfrac{{\bf p}^2}{2\mu} - {\cal E}_{\bf k} \right) 
\tilde{\psi}_{\bf k} ({\bf p}) 
=\dfrac{1}{4m_{t} \mu} 
\! \int \!
\dfrac{d^3 p'}{(2\pi)^3} 
{\cal K}^{\rm 2PI} (s,{\bf p,p'})
\tilde{\psi}_{\bf k} ({\bf p'}) ,
\label{eq:SchroedingerEq_MomentumSpace}
\end{align}
with ${\cal E}_{\bf k} = {\bf k}^2/(2\mu)$ being the kinetic energy of the system in the CM frame. Here, $m_{t}$ and $\mu$ stand for the total and reduced masses of the interacting particles. \Cref{eq:SchroedingerEq_MomentumSpace} can be Fourier transformed to position space using
\begin{align}
\psi_{\bf k} ({\bf r}) &= \int \dfrac{d^3p}{(2\pi)^3} 
\, e^{\im {\bf p \cdot r}}
\tilde{\psi}_{\bf k} ({\bf p}) ,
\label{eq:FT_wavefunction}
\\
{\cal V} ({\bf r,r'}) &= 
\dfrac{-1}{4m_{t} \mu} 
\!\int \!
\dfrac{d^3 p}{(2\pi)^3}  
\dfrac{d^3 p'}{(2\pi)^3} 
\, e^{\im {\bf p \cdot r}}
{\cal K}^{\rm 2PI} (s,{\bf p,p'})
\, e^{-\im {\bf p' \cdot r'}} ,
\label{eq:FT_potential}
\end{align}
where we left the dependence of ${\cal V} ({\bf r,r'})$ on $s$ implicit.

Two important remarks are in order. 
First, no on-shell condition should be imposed on the 2PI kernel, ${\cal K}^{\rm 2PI} (s,{\bf p,p'})$, since resummation involves its off-shell iterations of the 2PI diagrams. Note that when on-shell, $|{\bf p}|=|{\bf p'}|$ are determined by $s$. In fact, in the instantaneous approximation, the dependence of ${\cal K}^{\rm 2PI}$ on  the energy transfer is neglected (cf.~ref.~\cite{Petraki:2015hla}).  In position space this yields, in general, non-local potentials ${\cal V}({\bf r,r'})$.
Second, guided by the generalized optical theorem \eqref{eq:OpticalTheorem_gen}, we do \emph{not} restrict our analysis to the case of ${\cal K}^{\rm 2PI}$ depending on ${\bf p - p'}$ only, rather than ${\bf p}$ and ${\bf p'}$ separately. 2PI kernels that depend only on ${\bf p - p'}$ lead to central potentials, where all partial waves are related. We treat instead all partial waves independently, and do not specify the momentum dependence of ${\cal K}^{\rm 2PI}$.

In a weakly coupled system, the potential vanishes at $r\to \infty$ as ${\cal V}({\bf r,r'}) \sim r^{-(1+\epsilon)}$ with $\epsilon \geqslant 0$. Then, scattering is described by wavefunctions that asymptote to an incoming plane wave plus an outgoing spherical wave at spatial infinity; in position space
\begin{align}
\psi_{\bf k} ({\bf r}) \stackrel{r\to \infty}{\simeq} 
e^{\im {\bf k \cdot r}} + f_{\bf k} (\Omega_{\bf r}) \dfrac{e^{\im k r}}{r} ,
\label{eq:Wavefunction_AsymptoticForm}
\end{align}
where $k=|{\bf k}|$, and $f_{\bf k} (\Omega_{\bf r})$ encodes the scattering dynamics. (For ${\cal V}(r,r') \propto 1/r$, $f_{\bf k}$ has a mild $r$ dependence at $r\to \infty$, which does not spoil any of our conclusions.) 

As is customary, $\psi_{\bf k}$ and $f_{\bf k}$ may be expanded into partial waves and re-expressed in terms of phase-shifts, 
\begin{align}
\psi_{\bf k} ({\bf r}) 
&= \sum_\ell (2\ell+1) P_\ell ({\bf \hat{k}\cdot\hat{r}}) \, \psi_{|\bf k|,\ell} (r) ,
\label{eq:PartialWaveExpansion_WF}
\\
f_{\bf k} (\Omega_{\bf r}) 
&= \sum_\ell (2\ell+1) P_\ell ({\bf \hat{k}\cdot\hat{r}}) 
\, f_{|\bf k|,\ell} ,
\label{eq:PartialWaveExpansion_f}
\\
f_{|\bf k|,\ell}
&= 
\dfrac{e^{\im 2\Delta_{\ell}(k)}-1}{\im 2k} ,
\label{eq:f_AsymptoticForm}
\end{align}
where the phase shifts $\Delta_\ell (k)$ are in general complex. The partial-wave cross-sections are (see e.g.~\cite{Lifshitz_RelativisticQM})
\begin{subequations}
\label{eq:PhaseShifts}
\label[pluralequation]{eqs:PhaseShifts}
\begin{align}
\dfrac{\sigma^{\rm tot}_{\ell}}{\sigma^U_\ell} 
&= k \, {\rm Im} f_{|\bf k|,\ell}
= \dfrac{1}{2} \left[1 - {\rm Re} (e^{\im 2 \Delta_{\ell}}) \right],
\label{eq:PhaseShifts_sigmaTot}
\\
\dfrac{\sigma^{\rm elas}_{\ell}}{\sigma^U_\ell} 
&= k^2 |f_{|{\bf k}|,\ell}|^2 
= \dfrac{1}{4} \left[1+ e^{-4{\rm Im} \Delta_{\ell}} - 2{\rm Re} (e^{\im 2 \Delta_{\ell}}) \right],
\label{eq:PhaseShifts_sigmaElas}
\\
\dfrac{\sigma^{\rm inel}_{\ell}}{\sigma^U_\ell} 
&= \dfrac{\sigma^{\rm tot}_{\ell} - \sigma^{\rm elas}_{\ell}}{\sigma^U_\ell} 
= \dfrac{1}{4} \left[1- e^{-4{\rm Im} \Delta_{\ell}} \right].
\label{eq:PhaseShifts_sigmaInel}
\end{align}    
\end{subequations}
The total cross-section is obtained from the standard form of the optical theorem, and \cref{eq:PhaseShifts_sigmaInel} includes all inelastic channels.
The factors $2^\delta$ incorporated in $\sigma^U_\ell$ arise from the (anti)symmetrisation of the wavefunction in the case of identical particles that doubles the strength of the $\ell$ modes that survive; for all other $\ell$ modes, $\sigma_\ell$ vanish.

To ensure that the combined unitarity bound \eqref{eq:UnitarityBound_CrossSections_Combined} is respected by the above solution, we must still show that ${\rm Im} \Delta_\ell \geqslant 0$. This is related to the imaginary part of ${\cal K}^{\rm 2PI}$ that we discuss next. 

\section{Imaginary potential from generalized optical theorem \label{sec:ImaginaryPotential}}

The generalized optical theorem \eqref{eq:OpticalTheorem_gen} suggests that inelastic processes generate an imaginary ${\cal K}^{\rm 2PI}$ component. Resumming this contribution renders inelastic cross-sections consistent with unitarity, as we show next.  

An imaginary potential due to inelastic processes has been previously considered in Ref.~\citep{Blum:2016nrz}, for the purpose of unitarizing $s$-wave annihilation. However, the connection with the generalized optical theorem was not identified, resulting in ambiguity on the sign of the potential, which is an important issue. Moreover, a central potential was adopted, which does not allow to treat different partial waves independently~\citep{Yamaguchi:1954mp}, and the chosen potential, a $\delta$-function in position space, amounted to a specific choice for the underlying inelastic interaction. 
(See also \citep{Braaten:2017dwq} for a study of similar scope, and \citep{Aydemir:2012nz,Kamada:2022zwb}
for works that do not resum inelastic processes but propose related ans\"atze to improve perturbative amplitudes.)
Our treatment relies on resummation, handles all partial waves independently, and does not make model assumptions for the momentum dependence of the resummed process. The sole constraint is the generalized optical theorem~\eqref{eq:OpticalTheorem_gen}, as it should be. 

To obtain ${\rm Im}( {\cal K}^{\rm 2PI} )$, we must remove from the right-hand side of \cref{eq:OpticalTheorem_gen} all elastic contributions (which yield 2-particle-reducible diagrams), and consider only inelastic contributions with any 2PI factors amputated. We denote the latter as ${\cal A}^{{\rm inel},j}$. Note that ${\cal A}^{{\rm inel},j}$ is \emph{not} the full amplitude for the inelastic channel $j$; the latter requires resummation of the 2PI diagrams of the incoming particles (cf.~e.g.~\cite{Petraki:2015hla}). For simplicity, we will neglect contributions from $2\to n\geqslant 3$ scattering.

%%%%%%%%%%%%%%%%%%%%%%%%%
\begin{figure}[t!]
\centering
\includegraphics[width=0.9\linewidth] {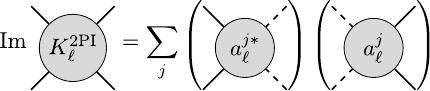}
\caption{The imaginary part of the 2PI kernel arises from squaring inelastic processes (cf.~\cref{eq:A2PI_Im}): the intermediate particles must be on-shell and different from the incoming / outgoing particles.
Here we show only 2-to-2 processes; multiparticle intermediate states require also integration over the allowed intermediate phase space. 
\label{fig:ResumFeynmanDiagrams_General}
}
\vspace{2em}
%%%%%%%%
\includegraphics[width=0.9\linewidth] {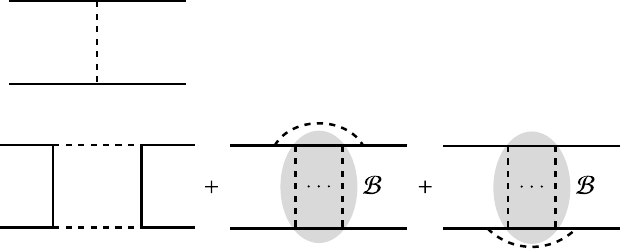}
\caption{
Examples of contributions to $K^{\rm 2PI}$ in the toy model discussed in \cref{sec:UnitarityLimitInterpretation}.  \emph{Top:} The one-boson exchange diagram is a  purely elastic interaction that generates the leading order contribution to ${\rm Re}(K^{\rm 2PI})$.
\emph{Bottom:} 2PI diagrams arising from squaring the 2-to-2 inelastic processes, 
$\chi\chi \to \varphi\varphi$ and 
$\chi\chi \to {\cal B}(\chi\chi)+\varphi$, where  ${\cal B}(\chi\chi)$ is bound state, described at leading order by an infinite ladder of one-boson exchange diagrams. These diagrams generate subleading contributions to ${\rm Re}(K^{\rm 2PI})$ from off-shell intermediate states, and leading contributions to ${\rm Im}(K^{\rm 2PI})$ from on-shell intermediate states. The latter are described by \cref{eq:A2PI_Im} (see text for explanations).
\label{fig:ResumFeynmanDiagrams_Examples}
}
\end{figure}
%%%%%%%%%%%%%%%%%%%%%%%%%

We partial-wave analyze ${\cal K}^{\rm 2PI}$ and ${\cal A}^{{\rm inel},j}$ according to \cref{eq:PartialWaveExpansion_Amplitude}, and define their corresponding rescaled versions, that we shall call $K^{\rm 2PI}_\ell$ and $a^j_\ell$, according to \cref{eq:Amplitudes_QMvsQFT_def}. Then, the generalized optical theorem,  \cref{eq:OpticalTheorem_gen,eq:OpticalTheorem_2-ParticleContribution}, imply
\begin{empheq}[box=\widefbox]{align} 
{\rm Im} [K^{\rm 2PI}_{\ell} (s,|{\bf k}|,|{\bf k'}|)] = 
\sum_{j} 
a^{j \, *}_\ell (s,|{\bf k}|) \, 
a^j_\ell (s,|{\bf k'|}) ,
\label{eq:A2PI_Im}
\end{empheq}
where we do not specify the final-state momenta in the inelastic factors $a^j_\ell$ since they are on-shell and determined by $s$. \Cref{eq:A2PI_Im} is a key element in our analysis. 
We depict it pictorially in \cref{fig:ResumFeynmanDiagrams_General}. 
Examples of contributions to $K^{\rm 2PI}$ from purely elastic and squared inelastic diagrams are contrasted in \cref{fig:ResumFeynmanDiagrams_Examples}.

To proceed, we transcribe \cref{eq:A2PI_Im} in position space, according to the Fourier transform \eqref{eq:FT_potential}. We analyze ${\cal V}({\bf r,r'})$ in partial waves,
\begin{align}
{\cal V}({\bf r,r'}) = \sum_{\ell} 
\dfrac{2\ell+1}{4\pi}
P_\ell({\bf \hat{r}\cdot\hat{r}'})
{\cal V}_{\ell} (r,r') ,    
\label{eq:PartialWaveExpansion_V}
\end{align}
and obtain
\begin{empheq}[box=\widefbox]{align} 
{\rm Im} [{\cal V}_\ell (r,r') ]
&=-\sum_j \nu^{j \, *}_\ell (r) \, \nu^j_{\ell}(r')  ,
\label{eq:V_Im}
\end{empheq}
with
\begin{align}
\nu^j_\ell (r) &\equiv \left( 
\dfrac{2^{1+\delta}}{\pi^2\mu} 
\right)^{1/2}
\int_0^\infty dp \, p^{3/2} \, j_\ell (pr) \, a^j_\ell (s,p) ,
\label{eq:nu_vs_a}
\end{align}
where we set $\sqrt{s} \simeq m_t$ in the prefactor, consistently with the non-relativistic approximation leading to \cref{eq:FT_potential}~\citep{Petraki:2015hla}.

We shall employ \cref{eq:V_Im} in \cref{sec:Regularization}, solve Schr\"odinger's equation and obtain the corrected wavefunction that, as we show, renders elastic and inelastic cross-sections consistent with unitarity. Before doing so, we offer here a simpler proof of this unitarization procedure, valid at leading order. 

We consider the current 
${\bf j}_{\bf k}({\bf r}) \equiv 
{\rm Im}     
[\psi_{\bf k}^*({\bf r}) \nabla \psi_{\bf k}({\bf r})]/\mu$. By virtue of Stokes' theorem and the continuity equation,
\begin{align}
\sigma^{\rm inel} 
&= -\dfrac{\mu}{k} 
\int d^3 r \, \nabla \cdot {\bf j}_{\bf k} ({\bf r}) 
\nonumber 
\\
&=
2\ {\rm Im} \left[ \int d^3 r  
d^3 r' 
\, \psi_{\bf k}^* ({\bf r})
{\cal V}({\bf r,r'}) 
\psi_{\bf k} ({\bf r'}) \right] .
\end{align}
Using the first Born approximation for the right-hand side, and expanding correspondingly the left-hand side in ${\rm Im}\Delta_\ell$ (cf.~\cref{eq:PhaseShifts_sigmaInel}), keeping lowest order terms, we find
\begin{align}
\sum_\ell \sigma^U_\ell {\rm Im} \Delta_\ell(k) \simeq 
\dfrac{1}{2m_t k}{\rm Im} [{\cal K}^{\rm 2PI} (s,k,k)] ,
\label{eq:Im_Delta}
\end{align}
which implies ${\rm Im}\Delta_\ell \geqslant 0$ as per \cref{eq:A2PI_Im}. (This remains so if we include $2\to n\geqslant3$ contributions to ${\rm Im}[{\cal K}^{\rm 2PI}]$.) 

The above analysis, distilled in \cref{eq:PhaseShifts,eq:A2PI_Im,eq:Im_Delta}, shows that the resummation of the imaginary part of the 2PI diagrams --- which arises from squaring inelastic processes, an essential point --- ensures that the elastic and inelastic cross-sections respect the combined unitarity bound \eqref{eq:UnitarityBound_CrossSections_Combined}. While this proof is at leading order in the couplings that give rise to inelastic scattering, and concerns the sum of all inelastic processes, we show in the following how the resummation of \cref{eq:A2PI_Im} ensures that unitarity is respected at all orders, and obtain unitarized solutions for the exclusive inelastic cross-sections.
 
\section{Unitarization \label{sec:Regularization}}

Next, we compute how the consistent resummation of the inelastic contributions to elastic scattering, and in particular of \cref{eq:A2PI_Im}, unitarizes the partial-wave elastic and exclusive inelastic processes.

To begin, we note that the elastic scattering cross-section will be computed according to \cref{eq:PhaseShifts_sigmaElas}, using the phase shifts that take into account the imaginary part of the potential given by \cref{eq:V_Im}. 
The cross-section for an inelastic channel $j$ is 
$\sigma^{{\rm inel},j}_{\ell} = \sigma^U_\ell |M^{{\rm inel},j}_{\ell}|^2$,  
where $M^{{\rm inel},j}_\ell$ stands for the corresponding amplitude rescaled according to \cref{eq:Amplitudes_QMvsQFT_def}, and incorporates both the hard-scattering inelastic amplitude, $a^j_\ell$, and the effect of the non-relativistic potential on the incoming particles, 
\begin{align}
M^{{\rm inel},j}_\ell (k) 
&=
\dfrac{k^{1/2}}{2\pi^2} 
\int_0^\infty dp 
\, p^{3/2} 
\, \tilde{\psi}_{|{\bf k}|,\ell} (p)
\, a^j_\ell (s,p)
\nonumber
\\
& =
\sqrt{\dfrac{2 \mu k}{2^\delta}}
\int_0^{\infty} dr\ r^2\ 
\psi_{|{\bf k}|,\ell} (r)
\nu^j_\ell (r).
\label{eq:Minel_def}
\end{align}
We must now determine the wavefunctions $\psi_{|{\bf k}|,\ell}$.

\subsection{Schr\"odinger's equation with complex potential}

We shall consider a complex potential whose real part we take to be central for simplicity, ${\rm Re}[{\cal V}({\bf r,r'})] = V(r) \delta^3({\bf r-r'})$, while its imaginary part is given by \cref{eq:V_Im}. Then, the partial-wave Schr\"odinger equation for 
$u_{|{\bf k}|,\ell} (r) \equiv k r\,\psi_{|{\bf k}|,\ell} (r)$ 
reads
\begin{multline}
{\cal S}_{\ell}(r)
u_{k,\ell} (r) -
\im 
\sum_j r \, \nu^{j*}_\ell (r) 
\int_0^\infty \!\! dr' r' \nu^{j}_{\ell} (r')
\, u_{k,\ell} (r') 
= {\cal E}_{\bf k} u_{k,\ell} (r),
\label{eq:SchroedingerEq_PW}
\end{multline}
where ${\cal S}_{\ell}(r)$ is the differential operator that includes the real part of the potential only,
\begin{align}
{\cal S}_{\ell}(r) \equiv 
-\dfrac{1}{2\mu}\dfrac{d^2}{dr^2} 
+ \dfrac{\ell(\ell+1)}{2\mu \, r^2} 
+V(r) ,
\label{eq:SchodingerOperator}
\end{align}
and ${\cal E}_{\bf k} = {\bf k^2}/(2\mu)$, as before.
We define two versions of the amplitudes and cross-sections: 
the \emph{unregulated} ones, for which we consider the real but neglect the imaginary part of the potential, 
and the \emph{regulated} ones, for which we take into account the entire complex potential. We denote them by the subscripts `unreg' and `reg' respectively, and aim to express the latter in terms of the former. 

To do so, we consider the 
solution of Schr\"odinger's equation in the former case and the Green's function,
\begin{align}
\left[{\cal S}_{\ell}(r) -{\cal E}_{\bf k}\right] 
{\cal F}_{k,\ell} (r) &= 0,
\label{eq:CentralPotential_Sol} \\
\left[{\cal S}_{\ell}(r) -{\cal E}_{\bf k}\right] 
{\cal G}_{k,\ell} (r) &= 0,
\label{eq:CentralPotential_Gsol} 
\\ 
\left[{\cal S}_{\ell}(r) -{\cal E}_{\bf k}\right] 
G_{k,\ell} (r,r') 
&= \delta(r-r'),
\label{eq:CentralPotential_GreensFun}
\end{align}
where ${\cal F}_{k,\ell}$ and ${\cal G}_{k,\ell}$ are the regular and irregular families of solutions that feature the following asymptotic behaviors at $r\to \infty$,
\begin{subequations}
\label{eq:CentraPotential_Sols_Asymptotes}
\label[pluralequation]{eqs:CentraPotential_Sols_Asymptotes}
\begin{align}
{\cal F}_{k,\ell} (r\to \infty) &\mapsto 
+\frac{1}{\im 2}
\left(
e^{\im kr} e^{\im 2\theta_\ell(k)}
- e^{-\im (kr-\ell \pi)}
\right) ,
\label{eq:CentraPotential_Fsol_Asymptote}
\\
{\cal G}_{k,\ell} (r\to \infty) &\mapsto 
-\frac{1}{2}
\left(
e^{\im kr} e^{\im 2\theta_\ell(k)}
+ e^{-\im (kr-\ell \pi)}
\right) .
\label{eq:CentraPotential_Gsol_Asymptote}
\end{align}
\end{subequations}
with $\theta_\ell (k) \in \mathbb{R}$. For some well-known real potentials, such as the Coulomb potential, ${\cal F}_{k,\ell}$ and ${\cal G}_{k,\ell}$ are known functions~\citep{Messiah:1962, Joachain:1975qct}. The functions ${\cal F}_{k,\ell}(r)$ describe states of an incident plane wave plus an outgoing spherical wave, and are the solutions of interest. ${\cal G}_{k,\ell}(r)$ will be useful for computational purposes. We shall require that $G_{k,\ell}(r,r')$ acts as an outgoing spherical wave at $r\to\infty$.  

We may now write an implicit solution of \cref{eq:SchroedingerEq_PW} for $u_{k, \ell}(r)$ for the full potential,
\begin{align}
\label{eq:SchroedingerWF_Ansatz}
&u_{k, \ell}(r) = 
{\cal F}_{k, \ell}(r) + 
\\
&+\im
\sqrt{\dfrac{2^\delta k}{2\mu}}
\sum_j
M_{\ell,\rm reg}^{{\rm inel},j}(k)
\left(
\int_0^\infty
dr'r'
G_{k, \ell}(r,r')
\nu_\ell^{j*}(r') 
\right) ,
\nonumber 
\end{align}
where we used
\begin{subequations}
\label{eq:Minel_RegUnreg_def}
\label[pluralequation]{eqs:Minel_RegUnreg_def}
\begin{align}
M^{{\rm inel},j}_{\ell, \rm reg} (k) 
&=
\sqrt{\dfrac{2\mu}{2^\delta k}}
\int_0^{\infty} dr \, r 
\, u_{k,\ell} (r)
\, \nu^j_\ell (r),
\label{eq:Minel_def_reg}\\
M^{{\rm inel},j}_{\ell, \rm unreg} (k) 
&=
\sqrt{\dfrac{2\mu}{2^\delta k}}
\int_0^{\infty} dr \, r 
\, {\cal F}_{k,\ell} (r)
\, \nu^j_\ell (r) ,
\label{eq:Minel_def_unreg}
\end{align}
\end{subequations}
to express the $r$-independent factor of the second term in terms of the rescaled regulated inelastic amplitudes, which depend on the wavefunction $u_{k,\ell}$. The unregulated amplitudes will become useful shortly.

Inserting \cref{eq:SchroedingerWF_Ansatz} into \eqref{eq:Minel_def_reg}, and introducing the matrix
\begin{align}
[\mathbb{N}_\ell (k)]^{ij}
\equiv
\delta^{ij}
-\im
\int_0^\infty
d r \, r 
\int_0^\infty
d r' \, r' \,
\nu_\ell^{i}(r)
\, G_{k,\ell}(r,r')
\, \nu_\ell^{j*}(r') ,
\label{eq:NormalMatrix}
\end{align}
we obtain a matrix equation between regulated and unregulated inelastic amplitudes. Upon inversion, it gives 
\begin{empheq}[box=\widefbox]{align} 
M^{{\rm inel},i}_{\ell, {\rm reg}} (k) =
\sum_j [\mathbb{N}_\ell^{-1}(k)]^{ij} \ 
M_{\ell,\rm unreg}^{{\rm inel},j} (k).
\label{eq:Minel_reg}
\end{empheq}
\Cref{eq:Minel_reg} is the key result of this work: it expresses  the regulated inelastic amplitudes in terms of the unregulated ones.  Similarly, using \cref{eq:Minel_reg}, the solution \eqref{eq:SchroedingerWF_Ansatz} can be re-expressed in terms of input parameters only: the imaginary potential and unregulated wavefunctions. 
\begin{empheq}[box=\widefbox]{align} 
\label{eq:Wavefunction_sol}
u_{k,\ell} (r)   =
{\cal F}_{k,\ell} (r) 
&+\im \sqrt{\dfrac{2^\delta k}{2\mu}}
\sum_{i,j} M_{\ell,\rm unreg}^{{\rm inel},j} (k)
[\mathbb{N}_\ell^{-1}(k)]^{ij}  \times
\nonumber
\\
& \times
\int_0^\infty dr' \, r' \, G_{k,\ell}(r,r') \nu_\ell^{i*}(r').
\end{empheq}

\subsection{Green's function}

To further simplify the results \eqref{eq:Minel_reg} and \eqref{eq:Wavefunction_sol}, we focus on the Green's function $G_{k,\ell} (r,r')$. We seek an expression for $G_{k,\ell} (r,r')$ that is regular at $r\to 0$ and has the asymptotic behavior of an outgoing spherical wave at $r\to \infty$.  Following Refs.~\citep{Yamaguchi:1954mp, McMillan:1963snp, Shah-Jahan:1973dlk, Joachain:1975qct, Arellano:2018ngv}, we expand the $r$ dependence of $G_{k,\ell}(r,r')$ in terms of ${\cal F}_{q,\ell}(r)$, which, being eigenfunctions of the hermitian operator ${\cal S}_\ell (r)$, constitute a complete orthonormal set of states with the desired asymptotic behavior,\footnote{\label{foot:NRapprox} This requires that the operator ${\cal S}_\ell (r)$ does not depend on the energy eigenvalue, which in turn implies that any dependence on $s$ has been eliminated by setting $s \simeq m_t^2$. Despite this non-relativistic approximation, the solutions to \cref{eq:CentralPotential_Sol,eq:CentralPotential_Gsol,eq:CentralPotential_GreensFun} can be defined for arbitrarily large energies. While such solutions may not well approximate the actual wavefunctions in the relativistic regime, they will be useful for computational purposes (cf.~\cref{foot:Analyticity}).
}
\begin{align}
G_{k,\ell} (r,r') 
&= \int_0^\infty dq \ g_{k,\ell} (q,r') \ {\cal F}_{q,\ell} (r) .
\label{eq:GreensFunction_expansion}
\end{align}
With the above decomposition, \cref{eq:CentralPotential_GreensFun} becomes
\begin{align}
\delta(r-r') 
&= \int_0^\infty dq 
\ g_{k,\ell} (q,r') 
[S_\ell (r) - {\cal E}_{\bf k}]
{\cal F}_{q,\ell} (r) ,
\nonumber \\
&= \int_0^\infty dq 
\ g_{k,\ell} (q,r') 
\left(\dfrac{q^2-k^2}{2\mu}\right)
{\cal F}_{q,\ell} (r) .
\label{eq:CentralPotential_GreensFun_expanded}
\end{align}
We must now invert the above to obtain $g_{k,\ell} (q,r')$. For this, we need the orthonormality relation for ${\cal F}_{q,\ell} (r)$, 
\begin{align}
\int_{0}^\infty dr \ 
{\cal F}_{q,\ell} (r)    
{\cal F}_{q',\ell}^* (r) =   
\dfrac{2^\delta \pi}{2} \delta(q-q').  
\label{eq:WavefunctionsNorm_PW}
\end{align}
Considering this, \cref{eq:CentralPotential_GreensFun_expanded} yields
\begin{align}
g_{k,\ell}(q,r) 
&= \dfrac{4\mu}{2^\delta \pi} 
\dfrac{{\cal F}_{q,\ell}^* (r)}{q^2-k^2} .
\label{eq:GreensFunction_expansion_Inverse}
\end{align}
Then, \cref{eq:GreensFunction_expansion} becomes
\begin{align}
G_{k,\ell} (r,r') = 
\dfrac{4\mu}{2^\delta \pi}
\int_0^{\infty} dq \,    
\dfrac{{\cal F}_{q,\ell} (r) {\cal F}_{q,\ell}^* (r')}
{q^2-k^2-\im \epsilon} .
\label{eq:SchroedingerEq_GreensFunc_Integral1}
\end{align}
In order to evaluate $G_{k,\ell} (r,r')$, it will be convenient to extend the range of integration over $dq$ to $(-\infty,\infty)$. To analytically continue to $q<0$, we consider the asymptotic behaviors \eqref{eq:CentraPotential_Sols_Asymptotes}, and set self-consistently
\begin{subequations}
\label{eq:AnalyticCont} 
\label[pluralequation]{eqs:AnalyticCont} 
\begin{align}
\theta_{\ell} (-q) &=-\theta_{\ell} (q) + \ell \pi, 
\label{eq:AnalyticCont_theta} 
\\   
{\cal F}_{-q,\ell} (r) &=-{\cal F}_{q,\ell}^* (r) ,
\label{eq:AnalyticCont_F} 
\\   
{\cal G}_{-q,\ell} (r) &=+{\cal G}_{q,\ell}^* (r)  .  
\label{eq:AnalyticCont_G} 
\end{align}
\end{subequations}
With this, \cref{eq:SchroedingerEq_GreensFunc_Integral1} becomes
\begin{empheq}[box=\widefbox]{align} 
G_{k,\ell} (r,r') = 
\dfrac{2\mu}{2^\delta \pi}
\int_{-\infty}^{\infty} dq \,    
\dfrac{{\cal F}_{q,\ell} (r) {\cal F}_{q,\ell}^* (r')}
{q^2-k^2-\im \epsilon} ,
\label{eq:SchroedingerEq_GreensFunc_Integral2}
\end{empheq}
with the prescription $\epsilon\to 0^+$ chosen such that $G_{k,\ell} (r,r')$ behaves as an outgoing spherical wave at $r\to\infty$, 
as will be shown below. Since ${\cal F}_{q,\ell}(r)$ and ${\cal G}_{q,\ell}(r)$ are real up to the same $\ell$- and $q$-dependent but $r$-independent phase, it follows that
\begin{subequations}
\label{eq:Reality} 
\label[pluralequation]{eqs:Reality} 
\begin{align}
{\cal F}_{q,\ell} (r) {\cal F}_{q,\ell}^* (r') 
&= {\cal F}_{q,\ell}^* (r) {\cal F}_{q,\ell} (r') ,
\label{eq:Reality_FF} 
\\
{\cal G}_{q,\ell} (r) {\cal G}_{q,\ell}^* (r') 
&= {\cal G}_{q,\ell}^* (r) {\cal G}_{q,\ell} (r') ,
\label{eq:Reality_GG} 
\\
{\cal F}_{q,\ell} (r) {\cal G}_{q,\ell}^* (r') 
&= {\cal F}_{q,\ell}^* (r) {\cal G}_{q,\ell} (r') .
\label{eq:Reality_FG}
\end{align}
\end{subequations}
Here, this implies $G_{k,\ell}(r,r') = G_{k,\ell}(r',r)$, as expected.

\smallskip

To evaluate \cref{eq:SchroedingerEq_GreensFunc_Integral2}, we consider the two independent eigenfunction families of ${\cal S}_\ell$, characterized by the asymptotic behaviors of \cref{eqs:CentraPotential_Sols_Asymptotes}. We then define
\begin{align}
{\cal H}_{k,\ell}^{(\pm)}(r) = {\cal F}_{k,\ell}(r)\pm \im {\cal G}_{k,\ell}(r) 
.
\label{eq:DefinitionInOutWavefunctions}
\end{align}
The functions ${\cal H}_{k,\ell}^{(\pm)}(r)$ have the properties
\begin{subequations}
\label{eq:SphericalWavesAsymptotic}
\label[pluralequation]{eqs:SphericalWavesAsymptotic}    
\begin{align}
{\cal H}_{k,\ell}^{(+)}(r\to\infty) 
&\mapsto - \im e^{+\im (kr + 2\theta_\ell)} ,
\label{eq:SphericalWavesAsymptotic_plus}
\\
{\cal H}_{k,\ell}^{(-)}(r\to\infty) 
&\mapsto + \im e^{-\im (kr - \ell \pi)} ,
\label{eq:SphericalWavesAsymptotic_minus}
\\
{\cal H}_{-k,\ell}^{(\pm)} (r) 
&=-{\cal H}_{k,\ell}^{(\pm)*} (r) .
\label{eq:AnalyticCont_H} 
\end{align}
\end{subequations}
We invert \cref{eq:DefinitionInOutWavefunctions} to express ${\cal F}_{k,\ell(r)}$ in terms of ${\cal H}_{k,\ell}^{(\pm)}(r)$, which allows to separate \cref{eq:SchroedingerEq_GreensFunc_Integral2} into two integrals,
\begin{align}
\label{eq:PositionSpaceGreensFunction_Contour}
G_{k,\ell} (r,r') 
&= 
\dfrac{\mu}{2^\delta \pi}
\int_{-\infty}^{\infty} dq \,    
\dfrac{{\cal F}_{q,\ell}^* (r) {\cal H}_{q,\ell}^{(+)} (r')}
{q^2-k^2-\im \epsilon}
\\
&+
\dfrac{\mu}{2^\delta \pi}
\int_{-\infty}^{\infty} dq \,    
\dfrac{{\cal F}_{q,\ell}^* (r) {\cal H}_{q,\ell}^{(-)} (r')}
{q^2-k^2-\im \epsilon} .
\nonumber
\end{align}
We evaluate the above via contour integration, using the residue theorem, and choosing the contours based on the asymptotic behavior of ${\cal H}_{k,\ell}^{(\pm)}(r)$. For $r < r'$, the first integral must be evaluated in the upper-half plane while the second in the lower half plane. The case $r > r'$ follows a similar logic. Considering also the analytic continuations \eqref{eq:AnalyticCont_F} and \eqref{eq:AnalyticCont_H}, as well as \cref{eqs:Reality}, we find, 
\begin{align}
G_{k,\ell}(r,r') = +\frac{2\mu \im}{2^\delta k} {\cal F}_{k,\ell}^*(r_<){\cal H}_{k,\ell}^{(+)}( r_>) ,
\label{eq:PositionSpaceGreensFunction}
\end{align}
where $r_<\equiv \min\{r,r'\}$ and $r_>\equiv \max\{r,r'\}$. The necessary asymptotic behaviour of $G_{k,\ell}(r,r')$ is made apparent by the appearance of ${\cal H}_{k,\ell}^{(+)}$ in the above expression.

If the spectrum of ${\cal S}_\ell$ includes bound states, they should be included in the Green's function spectral decomposition of \cref{eq:GreensFunction_expansion}. 
However, in this case, the scattering state wavefunctions ${\cal F}_{q,\ell}$ exhibit poles at the imaginary values of $q$ that yield the bound-state energies. Upon contour integration in
\cref{eq:PositionSpaceGreensFunction_Contour},
the residues of those poles cancel exactly the bound-state contributions from the spectral decomposition (cf.~e.g.~\cite{Mapleton:1961jmp}), thereby leaving the final result, \cref{eq:PositionSpaceGreensFunction}, unaffected.

\subsection{Regulated cross sections}
%%%%%%%%%%%%%%%%%%%%%%%%%
\begin{figure}[t!]
\centering
\includegraphics[width=0.975\linewidth] {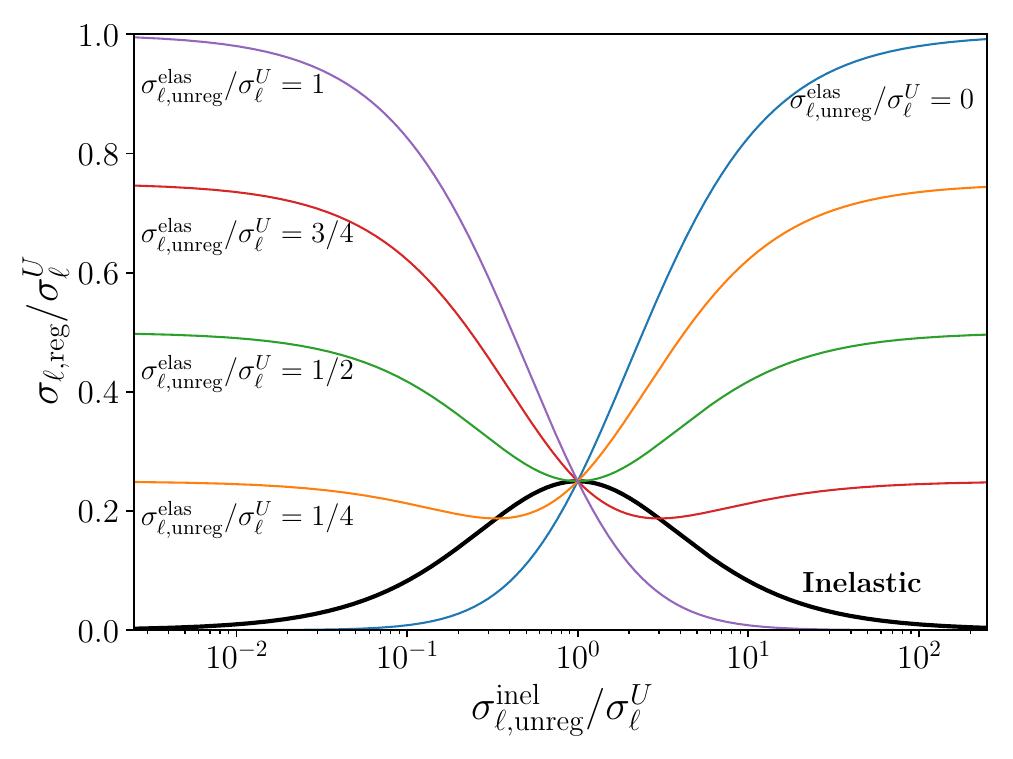}
\caption{The regulated elastic (colored lines) and 2-to-2 inclusive inelastic (black line) partial-wave cross-sections, vs the unregulated  2-to-2 inclusive inelastic cross-section.}
\label{fig:Unitarization}
\end{figure}
%%%%%%%%%%%%%%%%%%%%%%%%%

The results of the previous section can be used to derive simple expressions for the regulated elastic and inelastic cross sections in terms of the corresponding unregulated quantities. To begin, using \cref{eq:SchroedingerEq_GreensFunc_Integral2}, we can re-express the normalization matrix, \cref{eq:NormalMatrix}, as
\begin{empheq}[box=\widefbox]{align} 
[\mathbb{N}_\ell (k)]^{ij}
= \delta^{ij}
&- \frac{\im}{\pi} \int_{-\infty}^\infty
\frac{h_{\ell}^i(q) \, \bar{h}_{\ell}^{j}(q)}{q^2 - k^2 - \im\epsilon} \  d q ,
\label{eq:NormalMatrix_wrt_Munreg}
\end{empheq}
where, for $q\in \mathbb{R}$, we define
\begin{subequations}
\label{eq:h_defs}
\label[pluralequation]{eqs:h_defs}
\begin{align}
h_{\ell}^i(q) &\equiv \sqrt{\dfrac{2\mu}{2^\delta}} 
\int_0^\infty dr \, r \, {\cal F}_{q,\ell} (r) \nu_\ell^i (r),
\label{eq:h_def}
\\
\bar{h}_{\ell}^i(q) &\equiv 
{\rm Re} [h_{\ell}^i(q)] - \im~{\rm Im} [h_{\ell}^i(q)] .
\label{eq:hbar_def}
\end{align}
\end{subequations}
\Cref{eq:AnalyticCont_F,eq:Reality_FF} imply 
$h_\ell^i (-q) \bar{h}_\ell^{j} (-q) = h_\ell^i (q) \bar{h}_\ell^{j} (q)$.

With these definitions, the integrand in \cref{eq:NormalMatrix_wrt_Munreg} can be analytically continued in the complex $q$ plane, as it is an analytic function of $q$ except perhaps for isolated poles (meromorphic). If the functions $h_\ell^i(q)$ are holomorphic in the upper half of the complex plane and increase at complex infinity more slowly than $|q|^{1/2}$, then we can evaluate the contour integral to obtain,\footnote{\label{foot:Analyticity}
For physical, non-relativistic momenta, $0 < q \ll \mu$, the functions $h_\ell^i(q)$ are related to the unregulated inelastic amplitudes given by \cref{eq:Minel_def_unreg}, $h_\ell^i (q) = \sqrt{q} M_{\ell,{\rm unreg}}^{{\rm inel},i} (q)$. However, $M_{\ell,\rm unreg}^{{\rm inel},i} (q)$ cannot be analytically continued on the entire real $q$ axis, due to the rescaling by the non-analytic factor $\sqrt{q}$ relative to the full amplitudes in \cref{eq:Amplitudes_QMvsQFT_def}. In \cref{eq:NormalMatrix_wrt_Munreg}, we use $h_\ell^i(q)$ instead of $M_{\ell,\rm unreg}^{{\rm inel},i} (q)$ to make analyticity manifest. To ensure analyticity in the complex plane, we also use $\bar{h}_\ell^i (q)$, which depends only on $q$, instead of using $[h_\ell^i(q)]^* = \bar{h}_\ell^i(q^*)$, which introduces dependence on $q^*$. Finally, $h_\ell^i(q)$ is defined via \cref{eq:h_def} for arbitrarily large $|q|$, while \cref{eq:Minel_def_unreg} holds only for non-relativistic momenta (cf.~\cref{foot:NRapprox}). The large-momentum modes are needed for the spectral analysis of the Green's function, employed in \cref{eq:NormalMatrix_wrt_Munreg}.}
\begin{align}
[\mathbb{N}_\ell (k)]^{ij} =
\delta^{ij} +
M_{\ell,\rm unreg}^{{\rm inel},i} (k) 
M_{\ell,\rm unreg}^{{\rm inel},j*} (k).
\label{eq:N_analytic}
\end{align}
From here, it is easy to demonstrate that the amplitudes $M_{\ell,\rm unreg}^{{\rm inel},j} (k)$ form an eigenvector of $\mathbb{N}_\ell (k)$,
\begin{align}
\sum_{j}
[\mathbb{N}_\ell]^{ij}
M_{\ell,\rm unreg}^{{\rm inel},j} =
M_{\ell,\rm unreg}^{{\rm inel},i}
\left(
1 + \sigma^{{\rm inel}}_{\ell, {\rm unreg}} / \sigma^U_\ell
\right) .
\label{eq:Neigensystem}
\end{align}
The above implies that the regulated amplitude \eqref{eq:Minel_reg} is
\begin{align}
M_{\ell,\rm reg}^{{\rm inel},i} 
&=
\dfrac{M_{\ell,\rm unreg}^{{\rm inel},i}}
{1+ \sigma^{{\rm inel}}_{\ell, {\rm unreg}} / \sigma^U_\ell },
\label{eq:Ninverse_eigensystem}
\end{align}
from where which we obtain the regulated exclusive partial-wave inelastic cross-sections, which will be presented shortly.

Next, we want to determine the elastic cross-section. For this purpose, we need to deduce the complex phase shift 
$\Delta_\ell(k)$ that determines the asymptotic behavior 
of the partial-wave modes at $r\to \infty$, according to
\begin{align}
u_{k,\ell} (r\to\infty)
\mapsto
\dfrac{1}{\im 2} 
\left(
e^{\im [kr+2\Delta_{\ell}(k)]} - e^{-\im (kr-\ell \pi)} 
\right) .
\label{eq:Wavefunction_AsymptoticForm_PW}
\end{align}
We expand our solution, \cref{eq:SchroedingerWF_Ansatz}, in the $r\to \infty$ limit, using the Green's function \eqref{eq:PositionSpaceGreensFunction} and considering the asymptotic forms \eqref{eq:CentraPotential_Fsol_Asymptote} and \eqref{eq:SphericalWavesAsymptotic}. We find the asymptotic form \eqref{eq:Wavefunction_AsymptoticForm_PW} with the phase shift being
$\Delta_\ell = \theta_\ell + \delta_\ell$,
where
\begin{align}
e^{\im 2\delta_\ell (k)}
&= 1 
- 2 \sum_{i,j}
M_{\ell,\rm unreg}^{{\rm inel},i*} (k)
[\mathbb{N}_\ell^{-1}(k)]^{ij}
M_{\ell,\rm unreg}^{{\rm inel},j} (k)
\nonumber \\
&= 1
- 2 \left(
\frac{\sigma^{{\rm inel}}_{\ell, {\rm unreg}}/ \sigma^U_\ell }
{1+\sigma^{{\rm inel}}_{\ell, {\rm unreg}} / \sigma^U_\ell }
\right) ,
\label{eq:AdditionalPhaseShift}
\end{align}
where the second line has been obtained under the analyticity and convergence assumptions leading to \cref{{eq:N_analytic}}.
Note that ${\rm Re}(\delta_\ell) = 0$ or $\pi/2$ for
$\sigma^{{\rm inel}}_{\ell, {\rm unreg}} < \sigma^U_\ell$ or
$> \sigma^U_\ell$ respectively. From \cref{eq:AdditionalPhaseShift}, and given that $\sigma^{\rm tot}_{\ell} / \sigma_\ell^U = k\,{\rm Im}(f_{|{\bf k}|,\ell})$, or
\begin{subequations}
\label{eq:Elastic_PhaseShift}
\label[pluralequation]{eqs:Elastic_PhaseShift}
\begin{align}
\sigma^{{\rm elas}}_{\ell,\rm unreg} / \sigma_\ell^U
&= \dfrac{1}{2} \left( 1-{\rm Re} [e^{\im 2\theta_{\ell}(k)}] \right),
\\
(\sigma^{{\rm elas}}_{\ell,\rm reg}+\sigma^{{\rm inel}}_{\ell,\rm reg})/\sigma_\ell^U
&= \dfrac{1}{2} \left(1 - {\rm Re} [e^{\im 2\Delta_{\ell}(k)}] \right),
\end{align}
\end{subequations}
we can obtain the regulated elastic cross section. Defining
\begin{subequations}
\begin{align}
x_{\ell,\rm (un)reg} &\equiv
\sigma^{\rm elas}_{\ell, \rm (un)reg} / \sigma_\ell^U,\\[0.25cm]
y_{\ell,\rm (un)reg}^j &\equiv
\sigma^{{\rm inel},j}_{\ell, {\rm (un)reg}} / \sigma^U_\ell,\\[0.25cm]
y_{\ell,\rm (un)reg}^{} &\equiv \sum_j y_{\ell,\rm (un)reg}^j,
\end{align}
\end{subequations}
we find, using \cref{eq:Ninverse_eigensystem,eq:Elastic_PhaseShift},
\begin{subequations}
\label{eq:UnirarizedSigmas}
\label[pluralequation]{eqs:UnirarizedSigmas}
\begin{empheq}[box=\widefbox]{align} 
x_{\ell,\rm reg} &= \dfrac
{x_{\ell, \rm unreg} +(1-x_{\ell, \rm unreg}) \, y_{\ell,\rm unreg}^2}
{\left(1+ y_{\ell,\rm unreg} \right)^2} ,
\label{eq:UnirarizedSigma_Elas}
\\
y_{\ell,\rm reg}^j &= 
\dfrac{y_{\ell,\rm unreg}^j}
{\left(1+ y_{\ell,\rm unreg} \right)^2} .
\label{eq:UnirarizedSigma_Inel}
\end{empheq}
\end{subequations}
We reiterate that both the regulated and unregulated cross-sections include the effect of the real potential.

\Cref{eqs:UnirarizedSigmas} are manifestly consistent with unitarity: 
$x_{\ell,\rm reg} \leqslant 1$ and 
$y_{\ell,\rm reg}^j \leqslant y_{\ell,\rm reg} \leqslant 1/4$, and the combined bound \eqref{eq:UnitarityBound_CrossSections_Combined} is also respected. Note that $x_{\ell,\rm unreg} \leqslant 1$ since it already includes resummation of ${\rm Re}({\cal K}^{\rm 2PI})$. The maximum inelastic cross-section is reached for $y_{\ell,\rm unreg}=1$, where, consistently with \eqref{eq:UnitarityBound_CrossSections_Combined}, $x_{\ell,\rm reg}= y_{\ell,\rm reg}=1/4$, irrespectively of $x_{\ell,\rm unreg}$ (cf.~\cref{fig:Unitarity}). We plot \cref{eqs:UnirarizedSigmas} in \cref{fig:Unitarization}.

\section{Phenomenological implications}

The ramifications of the regularization \cref{eqs:UnirarizedSigmas} (or of the more general \cref{eq:Wavefunction_sol,eq:Minel_reg,eq:NormalMatrix_wrt_Munreg}) are far-reaching:
\begin{itemize}

\item 
The resummation of the squared inelastic diagrams affects both the inelastic and elastic cross-sections rather significantly. It is thus expected to affect the DM density in various production mechanisms, the DM indirect detection signals, and the DM self-interaction cross sections; the latter are important for the galactic structure (see \citep{Tulin:2017ara,Bullock:2017xww} for reviews).

\item 
Each inelastic channel is regulated by the \emph{inclusive} inelastic cross-section. 
It follows that even if a certain inelastic channel appears seemingly irrelevant for a particular purpose due to its nature, it may in fact be very important, provided that it is sufficiently strong. This is because such a process regulates the pertinent inelastic processes. 

For example, in DM freeze-out, metastable bound-state formation processes that are comparable to or faster than direct annihilation into radiation can significantly renormalize down the strength of the latter (cf.~Ref.~\cite[Fig.~11]{Oncala:2021tkz}), even when they themselves cannot deplete DM efficiently due to their rapid dissociation by the radiation bath. This may suppress the DM depletion and modify predictions.

Similarly, for DM indirect detection, rapid inelastic processes with (nearly) unobservable products can renormalize down inelastic processes that produce observable signals, thereby relaxing constraints.

\item 
For $y_{\ell,\rm unreg} \ll 1$, the relative corrections to the scattering rates are $\delta x_\ell /x_\ell \simeq \delta y_\ell^j /y_\ell^j \simeq -2y_{\ell,\rm unreg}$. 

For DM freeze-out, this implies that 
the imaginary potential affects the DM density at a level similar to or larger than the experimental uncertainty if $y_{\ell,\rm unreg} \gtrsim {\cal O}(10^{-2})$, i.e., even when the annihilation cross-section is far below the unitarity limit.  

\item 
At the opposite limit, $y_{\ell,\rm unreg} \gg 1$, we find 
$x_{\ell,\rm reg} \simeq 1-x_{\ell, \rm unreg}$ and 
$y_{\ell,\rm reg} \simeq 1/y_{\ell,\rm unreg}$, i.e., perhaps counter-intuitively, the regulated cross-sections decrease as the unregulated ones increase. (See below for an interpretation.)
Note that $y_{\ell,\rm unreg} \gg 1$ can occur even for (very) small couplings, e.g. at resonant points, as well as in bound-state formation processes where the potentials of the incoming scattering and outgoing bound state are different~\citep{Oncala:2019yvj,Oncala:2021swy,Oncala:2021tkz,Binder:2023ckj}. 
The small couplings ensure the validity of the instantaneous and non-relativistic approximations, and thus of the framework for the computations of this work.

With respect to the DM density, this implies that there could be two branches of solutions (mass-coupling correlation) that yield the observed value.

\item 
If $\sigma_{\ell,\rm unreg}^{\rm inel}$ scales with momentum in the same way as $\sigma_\ell^U$, then this feature is retained by the regularization. If not, the regularization generates a non-trivial velocity dependence. The earlier conclusion thus remains: the unitarity limits can be attained in a continuum of non-relativistic momenta only by long-range interactions. 

\item 
The upper bounds implied by unitarity on the mass of frozen-out DM annihilating via a single partial wave $\ell$, or all partial waves in the range $0\leqslant\ell\leqslant L$, are, for self-conjugate and non-self-conjugate DM,  
\\[6pt]
{\centering
\begin{tabular}{l|c|c}
$m_{\rm DM}^U / (140\ {\rm TeV})$
& solely $\ell$ 
& $0 \leqslant \ell \leqslant L$ 
\\ \hline
non-self-conj.~DM     
&  $\sqrt{1+2\ell}$
&  $(1+L)$
\\ \hline
self-conj.~DM     
& $2\sqrt{1+2\ell}$
& $2\sqrt{(1+L) \left(1+L/2 \right)}$
\end{tabular}
}\\[6pt]
where we used the numerical results of \citep{vonHarling:2014kha,Baldes:2017gzw}, assuming for definiteness that DM annihilates into plasma of the same temperature as that of the Standard Model, and the number of extra relativistic degrees of freedom is negligible.  It is straightforward to adjust these assumptions; this would affect the overall mass scale (here, 140~TeV), but not the scaling with $\ell$ or $L$. Several remarks are in order:

\begin{itemize}
\item 
The difference between the figures of 140~TeV denoted above, and 82~TeV deduced from \cite{Griest:1989wd} (cf.~\cref{sec:UnitarityLimitInterpretation}) is due to the momentum scaling of $\sigma_\ell^U$ that has been here properly taken into account~ \cite{vonHarling:2014kha,Baldes:2017gzw}.

\item
For self-conjugate DM, the above results incorporate the factor $2^\delta = 2$ of \cref{eq:sigmaU} that has not been previously appreciated in similar estimations. They also take into account that only $\ell$ even or odd modes survive, depending on the statistics. 

\item
For non-self-conjugate DM, the above takes into account only particle-antiparticle interactions, as is standard. However, DM may be depleted also via particle-particle inelastic scatterings (see e.g.~\cite{Oncala:2019yvj} for a model), which increases the bounds on the DM mass shown in the table by a factor $\sqrt{3}$. 

\item
We reiterate that there is no model-independent bound on the mass of thermal-relic DM due to unitarity, since which and how many partial waves contribute significantly depends on the model. 

\end{itemize}
\end{itemize}

We close this discussion with an interpretation of \cref{eqs:UnirarizedSigmas} in hydrodynamic terms. The resummation of inelastic interactions accounts for the reduction in the particle flux due to inelastic scatterings, as suggested by the continuity equation~\cite{Blum:2016nrz}. This reduction suppresses both elastic and inelastic scattering, by the factor $(1+y_{\ell,\rm unreg})^{-2}$. 

For inelastic scatterings, this effect competes against the fact that the probability for inelastic scattering increases with $y_{\ell,\rm unreg}$. For $y_{\ell,\rm unreg} \leqslant 1$, the increase of the inelastic scattering probability with  $y_{\ell,\rm unreg}$ dominates, while for $y_{\ell,\rm unreg} > 1$, the suppression of the particle flux takes over. 

Elastic scattering arises from two contributions. Particles may scatter due to the purely elastic couplings with probability $x_{\ell,\rm unreg}$. They evade this scattering with probability $1- x_{\ell,\rm unreg}$, but may scatter inelastically, with the products of the inelastic processes scattering back into the original species via the inverse reactions; this introduces an additional probability factor $y_{\ell,\rm unreg}^2$. The regenerated flux dominates elastic scattering at sufficiently large $y_{\ell,\rm unreg}$ (except if $x_{\ell,\rm unreg} =1$), where the regeneration probability $\propto y_{\ell, \rm unreg}^2$ exactly balances out the suppression $(1+y_{\ell, \rm unreg})^{-2}$, resulting in $x_{\ell,\rm reg} \simeq 1-x_{\ell,\rm unreg}$.

\section{Conclusions \label{sec:Conclusion}}

The violation of inelastic unitarity bounds in the non-relativistic regime by state-of-the art computations hampers a variety of phenomenological investigations in the frontiers of DM research. Resolving this issue is important for improving our theoretical understanding, as well as interpreting and guiding experimental searches. We have derived a simple analytical regularization scheme, given by \cref{eqs:UnirarizedSigmas}, whose input are the inelastic cross-sections as affected only by the real part of the potential, and output are the unitarized elastic and inelastic cross-sections.  The scheme applies to any partial wave, and is model-independent as it makes no assumptions about the momentum dependence of the unregulated inelastic amplitudes, except for analyticity and convergence requirements. (The more general \cref{eq:Wavefunction_sol,eq:Minel_reg} along with \cref{eq:NormalMatrix_wrt_Munreg} can be employed when these requirements do not hold.)
These results can be, and must be, employed in investigations of new physics scenarios in the non-relativistic regime, with wide-ranging implications.

\section*{Acknowledgements}
We thank Mark Goodsell, Miguel Paulos, Yifei He and Tobias Binder for useful discussions. This work was supported by the European Union’s Horizon 2020 research and innovation programme under grant agreement No 101002846, ERC CoG CosmoChart.

\appendix

\bibliographystyle{elsarticle-num}
\biboptions{sort&compress}
\bibliography{Bibliography}

\end{document}